\UseRawInputEncoding

\documentclass[preprint,10pt]{elsarticle}
\usepackage{amsmath}
\usepackage{amssymb}

\usepackage{color}

\let\mpar=\marginpar
\renewcommand\marginpar[1]{\mpar{\raggedright \scriptsize #1}}

\usepackage[labelfont=bf,labelsep=period,justification=raggedright]{caption}

\makeatletter
\renewcommand{\@biblabel}[1]{\quad#1.}
\makeatother
\date{}

\usepackage{graphicx}
\usepackage{dcolumn}%
\usepackage{bm}%
\usepackage{hyperref}

\def\be{\begin{equation}}   \def\ee{\end{equation}}

\usepackage{xr}
\begin{document}

\begin{flushleft}
{\Large
\textbf{Schr\"{o}dinger's ``What is Life?'' at 75}
}
\\
\bf{Rob Phillips}$^{\ast}$
\\
 Division of Biology and Biological Engineering and Department of
 Physics, California Institute of Technology, Pasadena, California, U.S.A
\\

$\ast$ E-mail: phillips@pboc.caltech.edu
\end{flushleft}

``These facts are easily the most interesting that science has revealed in our day.'' -Erwin 
Schr\"{o}dinger in ``What is Life?'',  discussing how the ``dislocation of just a few atoms'' in a gene can bring about
a ``well-defined change in the large-scale hereditary characteristics of the organism.''

\section*{Abstract}

2019 marked the 75th anniversary of the publication of  Erwin Schr\"{o}dinger's ``What is Life?'', a short book described by Roger Penrose in his preface to a reprint of this classic  as ``among the most influential scientific writings of
the 20th century.''  In this article, I review
the long argument made by Schr\"{o}dinger as he mused on how the laws of physics could help us understand  ``the events in space and time which take place within the spatial boundary of a living organism.''   Though Schr\"{o}dinger's book is often
hailed for its influence on some of the titans who founded molecular biology, 
this article takes a different tack.
Instead of exploring the way the book touched biologists such  as James Watson 
 and Francis Crick, as well as its critical reception by others such as Linus Pauling and 
Max Perutz, I argue that  Schr\"{o}dinger's classic
is a timeless manifesto,  rather than a dated
historical curiosity.    ``What is Life?'' is  full of timely outlooks and  approaches to understanding the mysterious living
world that includes and surrounds us and can instead be viewed as a call to arms to tackle the great unanswered challenges
in the study of living matter that remain for 21$^{st}$ century science.

\section{Background: The Author, His Book and its Readers}

\subsection{Schr\"{o}dinger and the circumstances of his public lectures}

Erwin Schr\"{o}dinger is one of the luminaries of twentieth-century physics.
His insights into the workings of the microscopic world were codified
in the famed wave equation that bears his name and gave rise (among many
other things) to
the s, p and d orbitals we all learn about in our first real encounter with
this world in high-school chemistry.  Schr\"{o}dinger was born and raised in Austria,
the only child of a highly intellectual family with a father deeply interested in
botany, leaving Schr\"{o}dinger himself with a sincere and enduring interest
in the living world.   In his  Autobiographical Sketches he recounts that because of
discussions with his father,  ``I had virtually devoured �The Origin of Species'' ...Of course I soon became an ardent follower of Darwinism (and still am today)~\cite{Schrodinger1992}.''

\begin{figure}
\centering{\includegraphics[width=3.0truein]{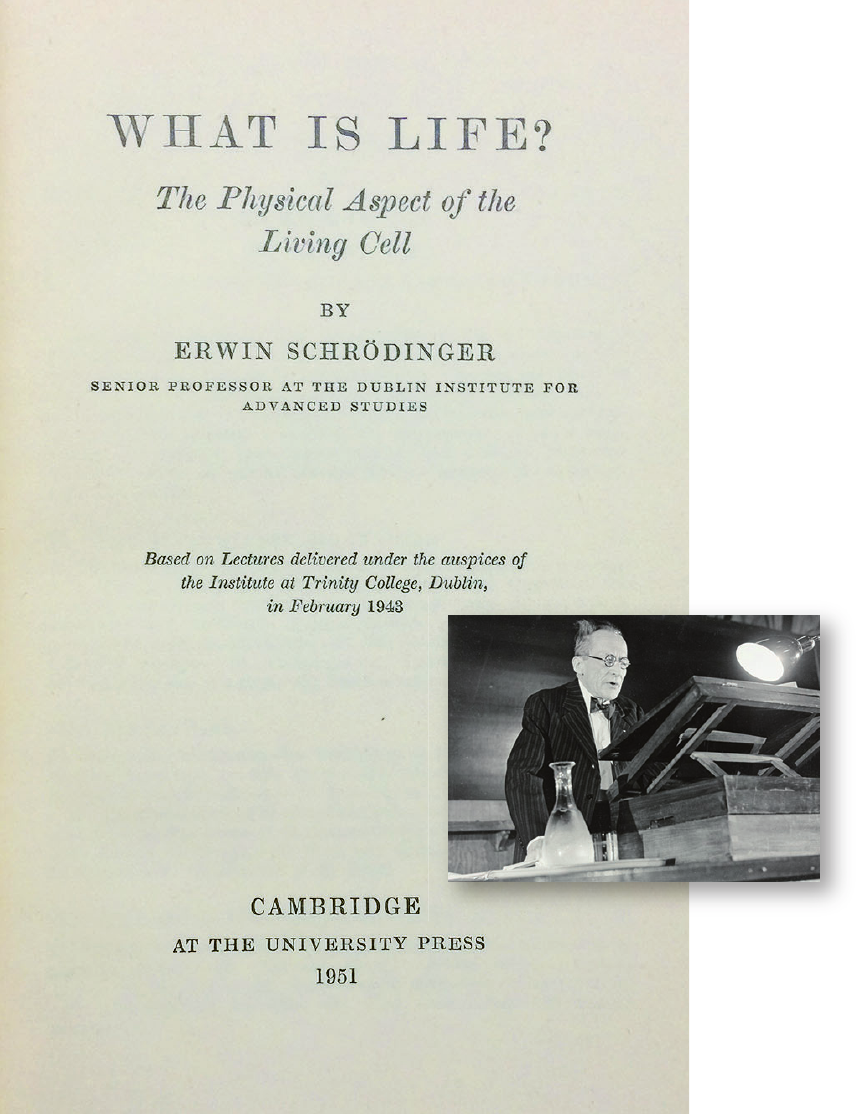}}
\caption{The 1944 edition of Erwin Schr\"{o}dinger's ``What is Life?,'' published
by Cambridge University Press.  The book resulted from a series of public lectures
offered by Schr\"{o}dinger in Dublin in 1943.
\label{fig:SchrodingerBook}}
\end{figure}

In the late 1930s after the {\it Anschluss}, Schr\"{o}dinger had to flee the Nazis
and turn his back on his professorship in Graz, Austria.  Fortunately, Irish prime minister \'{E}amon de Valera
was in the process of establishing the Institute for Advanced Studies in Dublin, and it was
 there that the great physicist  relocated.  Later,
speaking of his 17 years in Dublin, Schr\"{o}dinger called them
 ``the happiest years of my life,'' and it was in the context of his life there
 that he offered the public lectures that ultimately became the book we celebrate here
 (see  Figure~\ref{fig:SchrodingerBook} for the cover page
of the 1944 edition).
 To get a deeper sense of the impressive intellect that Schr\"{o}dinger brought to bear
 on topics far and wide such as his gift for languages (including English, Spanish and ancient
 Greek), the breadth of his accomplishments in physics and his personal lifestyle,
 there are several excellent sources~\cite{Dronamraju1999, Moore2015,Gribbin2013}.
Noted physicist Max Born's autobiography  gives us an impression of the regard Schr\"{o}dinger  commanded among his intellectual peers: ``His private life seemed strange to bourgeois people like ourselves.  But all this does not matter. He was a most lovable person, independent, amusing, temperamental, kind and generous, and he had a most perfect
 and efficient brain.''

\subsection{The Book's Mission: ``Accounting'' for Life}

Schr\"{o}dinger's ``What is Life?''  constitutes a long argument  in which he sets himself the task of nothing less than the search for ``unified, all-embracing knowledge'' to describe 
 the natural phenomenon we refer to as life.  In the section entitled ``The General Character and Purpose of the Investigation,'' he articulates his quest for universal knowledge more precisely by asking the oft-quoted question
``How can the events in {\it space and time} which take place within the spatial boundary of a living
organism be accounted for by physics and chemistry?''   And it is to answering that question
that the entirety of the long argument that follows  is devoted.  He asserts that
the intellectual ledger sheet was not yet balanced, with the data outpacing 
corresponding conceptual understanding that ``accounts'' for them, words that remain true to this day.
Indeed, to really put forth a critical reading of  Schr\"{o}dinger's text, we need to understand 
 what he
means by the word ``account,'' a point that was  overlooked by
many of his most ardent critics and which serves as the first  key point of this essay.

\subsection{Impact on the Founders of Modern Biology}

\begin{figure}
\centering{\includegraphics[width=4.3truein]{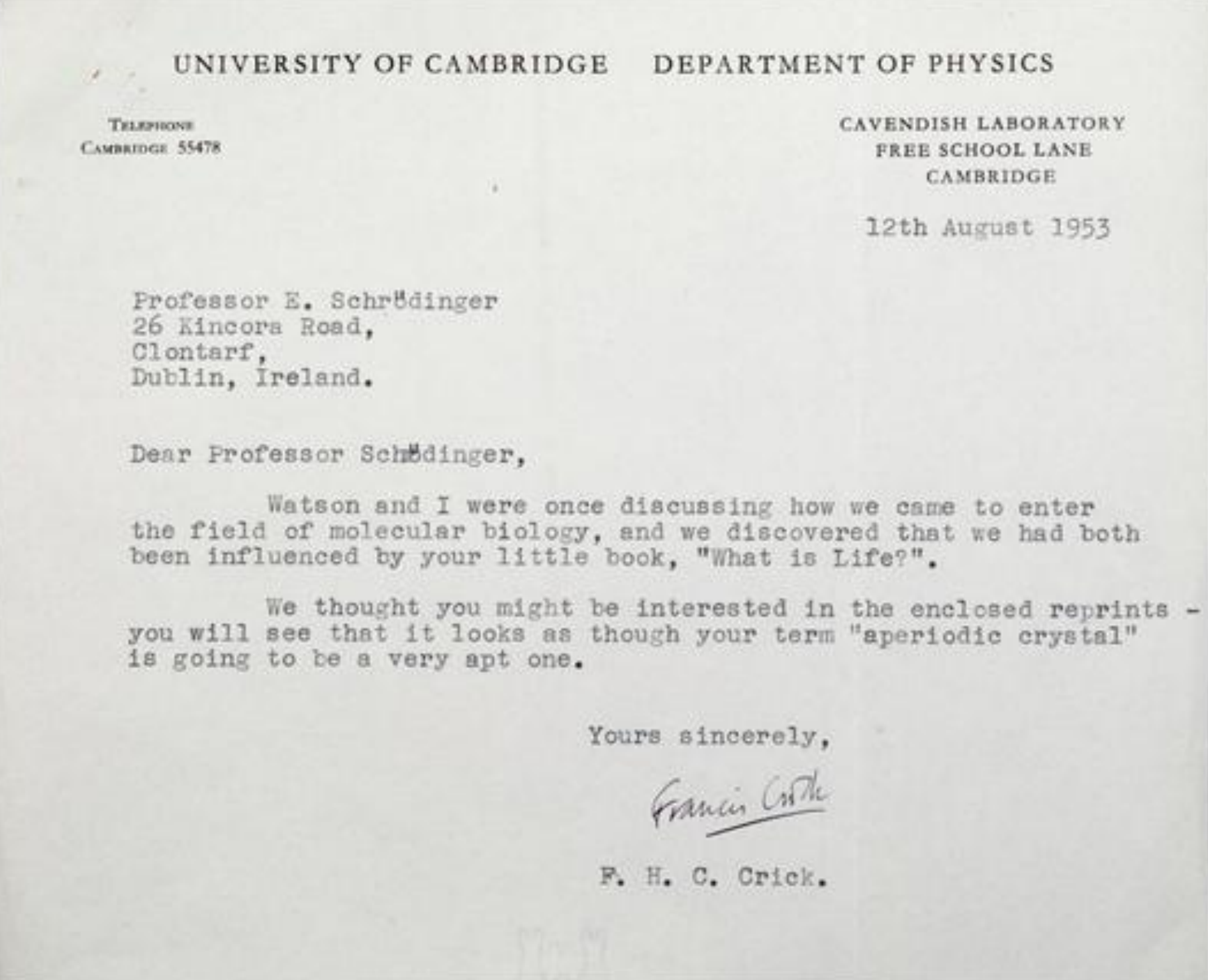}}
\caption{Letter from Crick and Watson to Schr\"{o}dinger openly acknowledging the importance
of ``What is Life?'' to their careers.
\label{fig:CrickWatsonLetter}}
\end{figure}

Most contemporary discussions of 
Schr\"{o}dinger's book focus not on its content but on its influence   on
a generation of biologists.  As a result, it seems important to touch briefly on
that story before turning to the main argument  that
Schr\"{o}dinger's call for a physics of the living world remains valid today.
Nearly a decade before the publication of the structure of DNA by Watson and Crick, 
Schr\"{o}dinger was already thinking about what came to be known as the coding problem.  As  Horace Freeland Judson
tells us in his classic history of modern biology, ``The Eighth Day of Creation''~\cite{Judson1996}:  ``The earliest mention of coding that counts was Erwin Schr\"{o}dinger's, in 
1944 in `What is Life?'  ...The fascination of the book lay in the clarity with which
Schr\"{o}dinger approached the gene not as an algebraic unit but as a physical substance that had to be
almost perfectly stable and yet express immense variety.''   
The influence of  Schr\"{o}dinger's book on some of the founders of modern biology  is evidenced by  
Figure~\ref{fig:CrickWatsonLetter}, which shows a letter from the young Francis Crick to the elderly master.

With the historical preliminaries  behind us, we turn instead to a different  question:  to what extent is 
Schr\"{o}dinger's classic relevant to the practice of biology and physics here and now?
I argue that  the vast majority of what Schr\"{o}dinger had to
say remains unfinished business.   This makes  it an opportune time at this three-quarters of
a century milestone to ask what would it
look like for the entirety of modern science, with no allegiance to any particular sub-discipline
within science, to revisit the question of the ``physical aspect of the living cell?''
The remainder of the essay centers on two key points: (i) Schr\"{o}dinger argues that the part of
the natural world we refer to as life needs to be ``accounted for'' in physical terms and (ii) the quest to
do such accounting will lead to new physics, not only helping us make sense of life, but also
enriching physics itself.

\section{The Meaning of ``Accounting for the Living Organism''}

Schr\"{o}dinger's charge was to find out how the actions within the walls of a living organism can
``be accounted for by physics and chemistry.''  When he uses the words ``account for'' what I think he had in mind was
the idea that our understanding of some phenomenon of interest can be seen as resulting from an appeal to some underlying principle, that that appeal is formulated in mathematical language in such a way that observed phenomena are seen
as quantitative consequences of these underlying principles and that the resulting insights make it possible
to make statements about phenomena not yet seen or measurements not yet made. 

\subsection{``Accounting For'' Inanimate Matter}

\begin{figure}
\centering{\includegraphics[width=6.0truein]{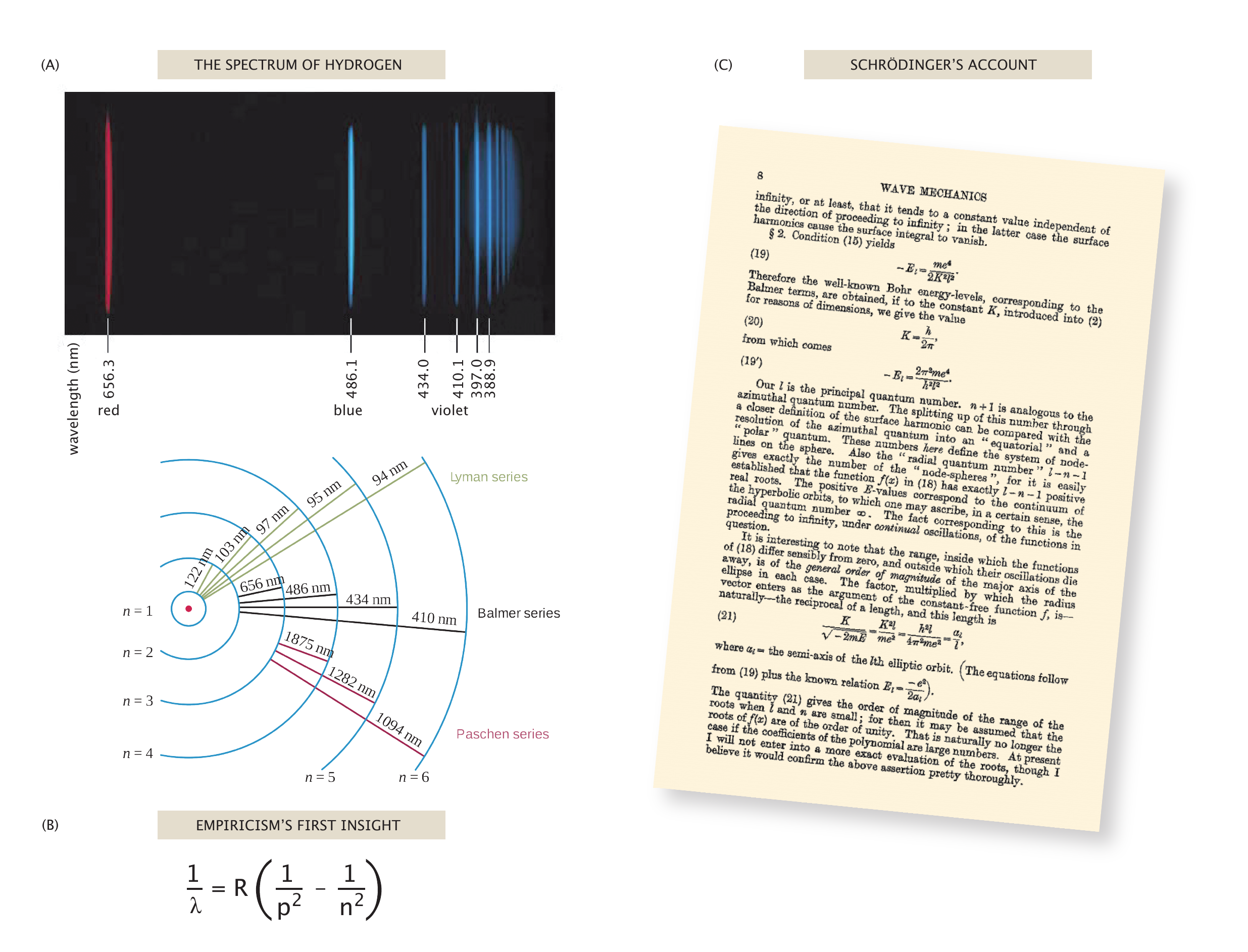}}
\caption{Accounting for the spectrum of hydrogen. (A) The spectrum of hydrogen with the wavelengths
of various spectral lines reported in nm.  (B) The Balmer formula for the wavelengths of
the spectral lines featuring the Rydberg constant, $R$.   (C)  Page from the English translation of Schr\"{o}dinger's paper on
wave mechanics in which he shows that the equation that now bears his name
amazingly gives rise to precisely the energy levels found earlier
by Niels Bohr that explain the spectrum of
hydrogen.
\label{fig:SchrodingerHydrogen}}
\end{figure}

A fitting analogy for what  Schr\"{o}dinger  meant
when talking of ``accounting for'' some class of phenomenon ``by physics and chemistry''  is provided
by his own work,
for which he is so deservedly famous.  In the mid- and late 19th century,
there was an explosion in our {\it factual} knowledge about the light
given off by and absorbed by different chemical elements in
the form of their atomic spectra as shown in Figure~\ref{fig:SchrodingerHydrogen}.  Just as with our current proliferation of gene and protein names and the burdensome nomenclature for the pathways
that connect them, the era of factual discovery in atomic spectra saw a proliferation of fascinating
and complicated spectral nomenclature with concepts such as the D-line of Na
and the different series such as the Balmer, Lyman, Brackett and Paschen series that are observed in the
H spectrum (for a sense of the enormity of the factual diversity of
these lines see~\cite{Harrison1939}).    But what set the wavelength of absorbed or emitted light for a given
element, or the number of spectral lines in a series?

Understanding in physics demands that factual knowledge based on
experiments be complemented by {\it conceptual}
knowledge.  Precisely the same kind of
unified, all-embracing knowledge that Schr\"{o}dinger  was speaking of in ``What is Life?'' was earlier
needed for thinking about the huge diversity of special cases found in atomic spectra
(for an enlightening description of the history of the study of atomic spectra,
see Mehra and Rechenberg~\cite{Mehra1982}). 
Nearly a century of struggle followed the discovery and experimental characterization of spectral lines as evidenced by the words of 
Max Planck in 1902:
``If the question concerning the nature of white light may thus be regarded as being solved, the answer to a closely related but no less important question - the question concerning
the nature of light of the spectral lines - seems to belong among the most difficult
and complicated problems which have ever been posed in optics or electrodynamics.''~\cite{Planck1902}
One outcome of these struggles (see Figure~\ref{fig:SchrodingerHydrogen})
was the discovery of empirical formulae that gave a phenomenological
mathematical description of the measured wavelengths of various spectral lines
seen in Figure~\ref{fig:SchrodingerHydrogen}(A)
of the form
\begin{equation}
{1 \over \lambda}= R\left( {1 \over p^2} - {1 \over n^2} \right),
\end{equation}
where $\lambda$ is the wavelength of the spectral line in question, $R$ is a phenomenological
parameter known as the Rydberg constant, later ``accounted for'' by physical understanding,
and $p$ and $n$ are integers.

To account for such phenomenological understanding requires us to construct theories
that explain why spectral lines are described by such a formula
and what determines the phenomenological constant $R$.
Niels Bohr's theory of the hydrogen atom gave a first tentative conceptual success
where he posited that the electron orbits are restricted to certain quantized values with energies 
\begin{equation}
E_n=- {Z^2 me^4 \over 8 h^2 \varepsilon_0^2 n^2},
\end{equation}
with $m$ the mass of the electron, $e$ the elementary charge of the electron,
$h$ the symbol for Planck's constant and $\varepsilon_0$ the permittivity of
free space.
Here the steps towards accounting take the form of the stunning realization that the empirical
Rydberg constant can in fact be written as
\begin{equation}
R= {me^4 \over 8 h^3 c \varepsilon_0^2} \approx 1.1 \times 10^7~\mbox{m}^{-1}.
\end{equation}
Erwin Schr\"{o}dinger's great atomic triumph, shown in Figure~\ref{fig:SchrodingerHydrogen},
was the insight that he could interpret the discrete energy levels of atomic
systems in much the same way that we interpret the musical notes from a guitar string
or an organ pipe using wave equations.  

Our digression into the niceties of how theoretical understanding ``accounted'' for 
the many complexities of spectral lines and so much more about the complicated
microscopic world is meant to serve as an invitation to
the kind of physical frameworks that Schr\"{o}dinger might have had in mind
when discussing living matter and using the words to ``account for'' a given phenomenon.  

\subsection{Accounting for Living Matter: From Thermal Energy to Shot Noise}

Though Schr\"{o}dinger clearly left the door open for quantum insights into
the phenomenon of life, much of his thinking centers on what is now known
as statistical physics.    In the 19$^{th}$ century, statistical physics arose 
in response to other phenomena demanding his type of accounting.
By then, mechanics had dominated as an
explanatory framework for more than a century, from Newton's 1687 presentation
of his ``System of the World,'' until the middle of the 19th century when questions
about the nature of heat began to take center stage~\cite{Klein1973, Brush1986}.  We can use  
Schr\"{o}dinger's language in that context as:  how can our experience of temperature
and heat  be accounted for by the mechanical motions of material
particles?   The quest to answer that question took the better part of a century,  culminating
in the twin edifices of statistical physics and thermodynamics, but also leaving
 in its wake unresolved problems such as the specific heats of crystalline solids
that would have to await the quantum theory for their proper resolution~\cite{Pais1979}.

Statistical physics was clearly much on Schr\"{o}dinger's mind in his Dublin years
 as evidenced not only by ``What is Life?,'' which is full of deep insights into 
 the subject that can be read with great profit with no reference to biology
 whatsoever, but also by his  1946 book,
 ``Statistical Thermodynamics,'' still a masterful treatise that one can learn from to
 this day.  Schr\"{o}dinger was schooled in the Austrian tradition of
Ludwig  Boltzmann's statistical physics, having only missed having one of the great founders (along with Maxwell and Gibbs) of
statistical mechanics as a professor by a year or two
due to Boltzmann's tragic  suicide.   But this did not stop  Schr\"{o}dinger from
being steeped in the tradition of statistical physics that ultimately became
central not only to our scientific understanding of classical physics, but also  to the way we view the quantum world he
helped uncover.   For Schr\"{o}dinger as announced in his Autobiographical Sketches, ``no perception in physics has ever seemed more important to me
than that of Boltzmann - despite Planck and Einstein,'' a resounding testament to the statistical physics sensibilities he brought
to the table in his thinking about the nature of life, and the shortcomings of which led to his belief in  the ultimate need for ``new physics'' to account for living matter~\cite{Schrodinger1992}.

As an introduction to the shortcomings of the physics of Schr\"{o}dinger's time to account for the living 
organism, Chapter 1 of ``What is Life?''  introduces ``The Classical Physicist's Approach to the Subject.'' 
Schr\"{o}dinger reminds us that statistical mechanics is the central conceptual
framework used until that time to interpret the collective properties of matter.   To that end, he begins by providing a beautiful and compelling introduction to the key
ideas of statistical physics.   Why?  Because, as he points out, ``it is in relation to the
statistical point of view that the structure of the vital parts of living organisms differs so entirely
from that of any piece of matter that we physicists and chemists have ever handled physically
in our laboratories or mentally at our writing desks.''  This idea of what Schr\"{o}dinger dubs
a ``difference in statistical structure'' between conventional and living matter is a theme that permeates the entirety of his short book.  The first chapter thus   lays the general groundwork for what
 follows by giving a highly simplified but profound view of statistical
physics.  There, he  raises many of the themes that dominate modern biology including the roles
played by
stochastic effects, small numbers, adaptation, accuracy and noise in the reproducible processes of
cells and organisms.

 In Chapters 2-5, he brings the general arguments
of Chapter 1 to bear on his first big specific question:   how  is genetic information
so stably passed from one generation to the next, despite 
how few atoms are  implicated in 
the gene (as we will discuss in the next section)?  Indeed, Schr\"{o}dinger argues that
the stability of the genetic material is inconsistent with the then-known laws of
statistical physics.
Though it is mind-boggling to consider,
one could say that in some sense biological information is  more stable than is the Earth itself.
In the 50 million or so years since the Indian subcontinent collided with Asia, 
8000 m mountains have been thrust up from the Earth, while at the same time,
the {\it Hox} genes that confer body plans in animals from flies to humans have
passed from one generation to the next such that the family resemblances are completely
evident when we compare sequences from different organisms.   By focusing on the nature of the hereditary molecule now known to be DNA, 
Schr\"{o}dinger foreshadowed
 the agenda for the DNA-centric perspective that colors much of modern biological investigation.

After concluding that the stability of genetic information leaves puzzles for
the physics of his (and our) day, Schr\"{o}dinger then turns to another set of biological
phenomena that  defy the physics of his time.  Here he focuses on what in modern guise we might call
the physico-chemical basis of the self-organization of living matter.
One of the most disturbing of human tendencies is
how we all become accustomed to the wonders around us,
whether in our science or in our lives.  One example that I find remarkable is seen on
transatlantic flights where, as we pass over the glaciers of Greenland, nearly
all passengers have their shades closed without even a thought of what lies
below.  
I might recast
Schr\"{o}dinger's question about the physico-chemical basis of the organization of
living cells thus:  how do the elements of
the periodic table, when sculpted by the consumption of energy at the nanometer scale through
ATP hydrolysis,  come together and form  remarkable
structures such as the insect eye or  the dynamic and constantly-renewing outer segment of animal photoreceptors?  These amazing
rhodopsin-filled structures  are  a reminder to do the scientific equivalent of
opening the shades over Greenland and marvel at    
living cells as collections of atoms from the periodic table that produce
``endless forms most beautiful and most wonderful.''

\begin{figure}
\centering{\includegraphics[width=5.0truein]{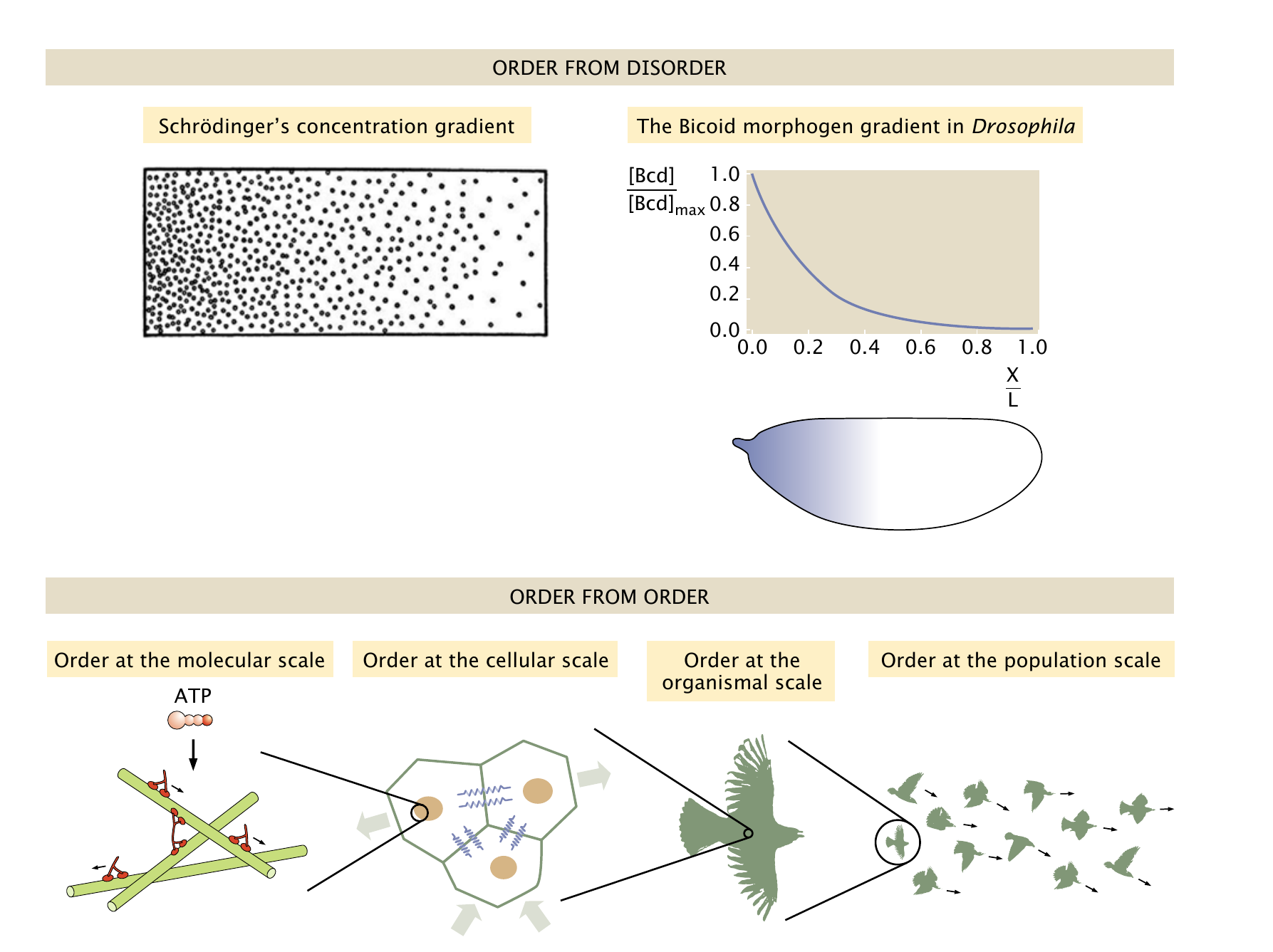}}
\caption{Schr\"{o}dinger's order from disorder and order from order.  In the top panel, Schr\"{o}dinger
illustrated the way that strictly random motions of atoms or molecules could give rise to reproducible patterns of concentration (left).  Modern developmental biology has borne that out in the form of morphogen  gradients such as the gradient of the transcription factor Bicoid in the fly embryo shown here schematically beneath the graph. The length ($L$) of the embryo is plotted
in dimensionless units $x/L$.  The profile of Bicoid can be thought of
as emerging from the ``disordered'' processes of diffusion and protein degradation. In the bottom panel,
the expenditure of energy at very small scales in the form of ATP hydrolysis leads to order at the
cellular scale, the organismal scale and even the population scale. 
\label{fig:NegativeEntropy}}
\end{figure}

Let's look a little more closely at the way that Schr\"{o}dinger examined the
capacity of statistical physics to account (or not) for living matter.
As a first example of the statistical structure of classical physics, he 
tackles  the question of paramagnetism (how  the tiny
individual magnets of single atoms conspire to give rise to the macroscopic
phenomenon  of magnetism and how it depends upon temperature).
The magnetic phenomenon results from a competition between the energy advantage that comes from aligning spins with
an applied magnetic field and the entropy that comes from those spins adopting random orientations. 
  The key point is the recognition
that the thermal energy scale, $k_BT \approx 4.1$~pN nm, as Schr\"{o}dinger points out, is the natural
energy scale that presides over the atomic and molecular world (and the molecules
of the living world can only find immunity from this energy scale by investing
some other energy such as ATP hydrolysis to avoid it - see the bottom panel of  Figure~\ref{fig:NegativeEntropy}).

With the ``order from disorder`` exhibited by magnetic materials in hand,
he then proceeds to a  description of Brownian motion.
   Schr\"{o}dinger's little four page
discussion of diffusion culminating in the diffusion equation itself  is as perfect a description of the subject as I have ever seen.   The important point of this discussion is to illustrate
the lawful mathematical precision  that emerges from the apparent lawlessness of the flips of
a coin as each molecule randomly chooses to move left or right, up or down. Speaking of the 
jostling of these Brownian particles, Schr\"{o}dinger poetically
muses:
``Their movements are determined by the thermic whims of the surrounding medium; they have no choice.  If they had some locomotion of their own, they  might nevertheless succeed in getting from
one place to another - but with some difficulty, since the heat motion tosses them like a small boat
in a rough sea.''   Again, the statistical structure of classical physics reveals that  ordered
states with the mathematical precision of an exponential function such as the concentration gradient shown in Figure~\ref{fig:NegativeEntropy} in fact
emerge from the second law of thermodynamics which tells us that systems evolve towards
states of maximum entropy, yielding a kind of order from disorder.   Turing's paper
showing an even more subtle and beautiful example of order from disorder was still a decade in the future~\cite{Turing1952}.

Once these
foundations have been laid, Schr\"{o}dinger  starts homing in on biological phenomena by considering the ``limits of accuracy of measuring,'' the insight being that there are limits to such measurement due to the natural fluctuations
of the measuring device itself, a phenomenon perhaps even more important  in living organisms
than in the experimenter's apparatus.
  Indeed, in my view, in this section,
Schr\"{o}dinger foreshadowed one of the most important themes of modern physical biology, 
namely, how living organisms defy the strictures of equilibrium physics.
These questions arise in settings ranging from
the origins of
high fidelity polymerization and the emergence of the concept of kinetic proofreading~\cite{Hopfield1974,Ninio1975} to the measurement of concentration differences
by cells performing chemotaxis~\cite{Berg1977} to the emergence of herds with orientational order in apparent defiance of
fundamental physical theorems~\cite{Toner2018}.
The topic of fidelity in biological polymerization was  also brilliantly undertaken by Linus Pauling~\cite{Pauling1957}, but from
a molecular perspective  rather than the conceptual point of view adopted by  Schr\"{o}dinger.

Schr\"{o}dinger starts his  discussion on the limits of accuracy of measuring with an analysis of the kinds of   torsional balance apparatus
used by Cavendish (gravity) and Coulomb (electrostatics) to measure the classic
inverse-square laws.  His point is that as the apparatus gets smaller and smaller,
the deflections induced by the forces of interest will be comparable in magnitude
to those induced by thermal motion.  
Questions of biological accuracy and precision have now
become a centerpiece of rigorous thinking in modern biology in the form of
the Berg-Purcell limit, but more generally in the context of how well living organisms can sense
their environments, whether in the context of hearing or seeing, or in the detection of
chemical messengers~\cite{Bialek2012}.  Schr\"{o}dinger himself understood precisely
the physics question in play, noting: ``The uncontrollable effect of the heat motion competes
with the effect of the force to be measured and makes the single deflection observed insignificant.
You have to multiply observations, in order to eliminate the effect of the Brownian movement of
your instrument.  This example is, I think, particularly illuminating in our present investigation.
For our organs of sense, after all, are a kind of instrument.  We can see how useless they would
be if they became too sensitive.''

But this is all more than pretty words.  Schr\"{o}dinger wants to describe these
physical challenges to the living organism quantitatively.  He notes the monotonous repetition of
this same statistical principle in the inorganic world,
but wants his listeners and readers to know that there is a simple mathematical
law, the ``so-called $\sqrt{n}$ law,'' that tells us the size of the fluctuations to
be expected in a system containing $n$ atoms or molecules.  
Schr\"{o}dinger summarizes that law as:
 ``The laws of physics and physical
chemistry are inaccurate within a probable relative error of the order of
$1/\sqrt{n}$, where $n$ is the number of molecules that co-operate to bring about
the law - to produce its validity with such regions of space or time (or both) that matter,
for some considerations or for some particular experiment.''  
These ideas
are central to modern biological enquiry.   For example, when photosynthetic bacteria
divide, they have to carry with them copies of a photosynthetic organelle known as the
carboxysome, with only a few (3-6) copies per cell.  During the division process, they have
special machinery to prevent these unwanted  $1/\sqrt{n}$ partitioning errors~\cite{Savage2010}.  By way of contrast, partitioning errors in transcription factors are a demonstrable part of the
reason for noisy gene expression~\cite{Rosenfeld2005, Rosenfeld2006}.
And yet, and here is the crux of
the whole book,  ``The classical physicist's expectation, far from being trivial, is wrong.''   As seen by
the carboxysome example, living organisms
have found ways to insulate themselves from the all-important $\sqrt{n}$ law and many of the other strictures of
equilibrium statistical physics.  

As I will elaborate on
below, one way to think of that ``wrongness'' is summarized in the bottom panel of Figure~\ref{fig:NegativeEntropy} which hints at the need for a statistical physics of the phenomena that emerge when energy is invested to keep those systems out of equilibrium, an enterprise that has been undertaken to great effect in recent decades (for a flavor of this new physics, see~\cite{Seifert2012}).
To further elaborate on why the classical physicist's expectation is wrong,  let's revisit Schr\"{o}dinger's quantitative analysis of the question of the stability of the genetic information.

\section{Fermi Problems in Schr\"{o}dinger Style}

Numbers sharpen our questions. 
Schr\"{o}dinger appreciated this sharpness and used quantitative estimates   as a key part of his arguments.
In the world of physics, numerical estimates to describe some
phenomenon of interest are sometimes known as Fermi problems and are a model for the power
of order-of-magnitude reasoning  to clarify both our
questions and our thinking in response to those questions.  Their name refers to a style of thinking brought to an art  form  by Italo-American physicist Enrico Fermi, who could  estimate his way to
numerical answers to questions ranging from the number of piano  tuners in a big
American city, to  the heat loss in our homes if  we forget to install storm windows,
to the yield of the atomic explosion in the Trinity Test of 1945.
Fermi brought this same  approach~\cite{Mahajan2010} to critical questions in the science challenges he faced in his professional life as a physicist.  
For example, 
the design of the first successful nuclear reactor - built  upon a string of systematic preliminary studies,
each characterized by numerical estimates on topics  such as neutron diffusion - led
Fermi to  predict   the precise moment
that the Stagg Field nuclear reactor would reach a self-sustained nuclear reaction~\cite{SchwartzBook}.

Schr\"{o}dinger's book draws from the same Fermi-problem inkwell.
In fact, Schr\"{o}dinger's short manifesto gives perhaps the best example
of a physicist working through Fermi problems in plain view that I have 
seen in written form. Over and over, Schr\"{o}dinger poses questions of the form: what sets the scale of X?, the quintessential framing of questions in order-of-magnitude
thinking.   Examples include: what sets the relative scale of organisms and atoms, what is the size of a gene
and how much energy is needed to maintain the
stability of the gene?

Having laid down his
 statistical mechanical and order-of-magnitude thinking foundations,   Schr\"{o}dinger was now prepared to take up
 the question of the size of a gene.
  His musing on this question is motivated by
  the statistical structure of classical physics and his interest
 in how biological systems deal with  the  $\sqrt{n}$ law introduced above.   His argument is that if the number of atoms associated with a gene is
small, then the natural fluctuations of such systems would  impact the stability of
genetic information.
Concretely, he wonders, how many atoms are associated with a gene
and if that number is small in the sense of the $\sqrt{n}$ law, then how do these genes
safeguard themselves from the inevitable statistical fluctuations present in any collection of
atoms? 

\subsection{On the Physical Dimensions of Genes}

\begin{figure}
\centering{\includegraphics[width=5.2truein]{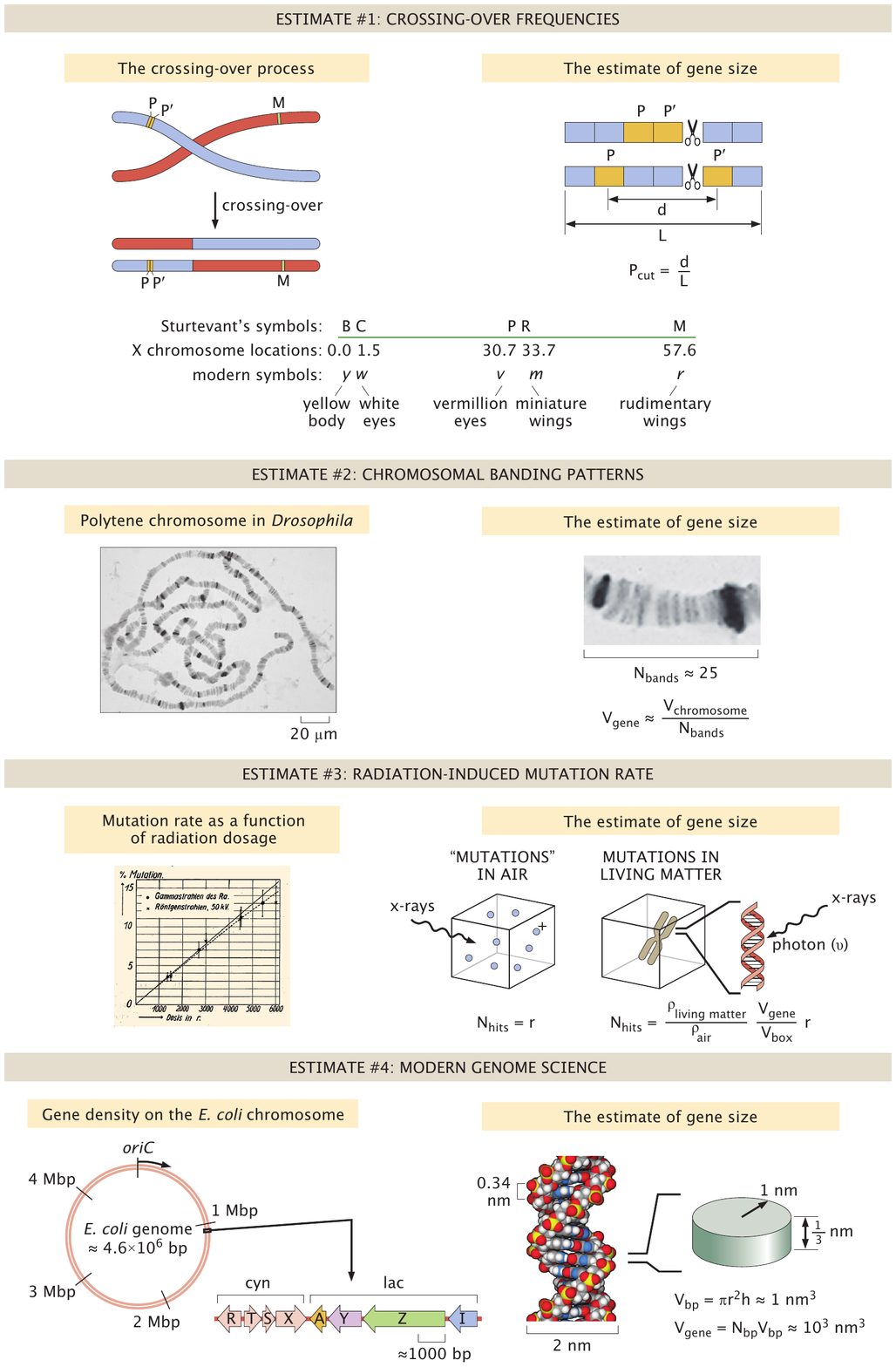}}
\caption{The size of a gene.  The first estimate exploited the beautiful insights into
mapping genes on chromosomes (originally in {\it Drosophila}) to figure out the mean spacing
of genes.   Given the gene number and the size of chromosomes this approach would yield helpful insights.   Schr\"{o}dinger's second estimate   uses banding patterns on chromosomes on  the mistaken assumption
that each band is a gene.    The Delbr\"{u}ck estimate was based on the mutation rate induced by x-rays. 
A modern estimate uses the number of base pairs in a gene and the very useful rule
that the volume of a base pair is 1~nm$^3$.  For more details of the paper of Timof\'{e}eff-Ressovsky, Zimmer and Delbr\"{u}ck that measured the data shown on mutation rate and radiation dosage, see the book ``Creating a Physical Biology'' which includes its translation~\cite{Sloan2011}.
\label{fig:GeneSize}}
\end{figure}

To examine the question of the size of a gene, Schr\"{o}dinger
 performs several different estimates 
which are illustrated in Figure~\ref{fig:GeneSize}.  Note the 
painstaking and often misguided detective work involved in answering such a deep question as the size of
the gene  before it was even widely accepted that DNA is the genetic material. 
 Schr\"{o}dinger begins with
two entirely independent estimates of the size of the gene.    The first,  shown 
in Figure~\ref{fig:GeneSize}(A) builds on
the subsection of his book entitled ``Crossing-over. Location of properties'' in which he explains how
maps like the classic case for {\it Drosophila} of Alfred Sturtevant are
discovered.  
As seen in
the figure, when there is a crossing over event, we can ask the frequency with which
two genes end up on the same strand together.  The more likely that occurrence, the closer
those two genes are on the chromosome and from these frequencies an actual map
of gene positions on the chromosome can be divined.
Given one of these maps of a  chromosome, Schr\"{o}dinger proposes
a strictly order-of-magnitude upper bound on the size of a gene based on the size of a chromosome and the
number of ``properties'' (i.e., genes) per each chromosome.  
As he notes, the estimate provides a bound,
because at the time of his writing, the entirety of the genes on a chromosome had not yet
been mapped.  The concept of the estimate is reasonable as is indicated for a bacterial
genome in Figure~\ref{fig:GeneSize}(D).
The second such estimate is a very clever
(though ``wrong'') idea of using the banded patterns of chromosomes as a measure of
gene size.  As seen in Figure~\ref{fig:GeneSize}(B), if we consider the 23 chromosomes (Schr\"{o}dinger refers to 24 pairs of chromosomes, true for chimps, but not humans) with their banding
patterns, this  estimate would say that the size of a gene is much larger than we know to be true.

He really hits his stride when considering a third estimate of the size of
a gene as seen in Figure~\ref{fig:GeneSize}(C).  This is the point in the story
where Schr\"{o}dinger's reports on the work of  Max Delbr\"{u}ck and coworkers, inspiring Perutz's remark
``In retrospect, the chief merit of ``What is Life?'' is its popularization of the Timof\'{e}eff, Zimmer and Delbr\"{u}ck paper that would otherwise have remained unknown outside the circles of geneticists and radiation biologists"~\cite{Perutz1987}. 
 As seen in the figure,
experiments on ionization of gases allowed for a measure of radiation intensity that
could then be directly translated into a tool for examining radiation damage in DNA.  
By measuring
the radiation-induced mutation rate, an estimate for the size of a gene could be made leading
to the idea that a gene involves roughly 1000 atoms.  Our modern understanding of the DNA double helix tells us that a base has $few \times 10$ atoms, meaning that a typical thousand nucleotide bacterial gene would involve
on the order of $few \times 10^4$ atoms, a bit more than a factor of 10 larger than Schr\"{o}dinger's estimate.  Though his estimate is too low for
the size of an entire gene and a little too large for the size of a basepair, it is 
an impressive use of indirect reasoning to arrive at molecular
dimensions that conjures images of the way Benjamin Franklin, Lord Rayleigh, Agnes Pockels
and Irving Langmuir
attacked the question of the molecular dimensions of lipid molecules indirectly by looking at the spreading
of oil on water~\cite{Tanford2004, Langmuir1917}.

The outcome of these arguments leaves Schr\"{o}dinger with a mystery:
``We are now seriously faced with the question: How can we, from the point of
view of statistical physics, reconcile the facts that the gene structure seems to involve
only a comparatively small number of atoms (of the order of 1000 and possibly much
less), and that nevertheless it displays a most regular and lawful activity - with a durability
or permanence that borders upon the miraculous?''  In examining this question, we have to put ourselves in the mindset of Schr\"{o}dinger  who is once again asking us from the perspective
of what it means for {\it physics} (not biology) to account for these phenomena.
Schr\"{o}dinger then goes farther by noting
``Thus we have come to the conclusion that an organism and all the biologically
relevant processes that it experiences must have an extremely `many-atomic' structure
and must be safeguarded against haphazard, `single-atomic' events attaining too
great an importance.''  
All of this kind of  ``what sets the scale of X'' thinking brought Schr\"{o}dinger to precisely
the same place that was arrived at 30 years later in a different way by John Hopfield
and Jacques Ninio when they presented a theory of just the kind of ``safeguarding'' called
for in ``What is Life?'' and that now goes under the name of kinetic proofreading~\cite{Hopfield1974,Ninio1975}.

\section{The Biological Frontiers of Physics: New Laws to Be Expected in the Organism}

Having made his arguments about the shortcomings of the statistical physics of his time
to account for the stability of living matter, Schr\"{o}dinger closes the book with some higher-level thinking 
on the implications of living matter for the future of physics.  In their
great book ``The Evolution of Physics,'' Albert Einstein and Leopold Infeld make it
very clear how physics has repeatedly been driven forward by new experimental measurements
(think Faraday and electromagnetic induction) that demand the emergence of
new concepts~\cite{Einstein1938}. 
 At the beginning of the 19$^{th}$ century, the concept of
entropy, one of the foundational ideas of modern science, had not even been
conceived or defined, but an increasingly sophisticated experimental program
revealing the character of temperature and heat demanded it.  Similarly, although the idea of ``field theory'' was implicit in
the development of continuum mechanics by great thinkers such as Euler, it would
have to await the labors of Faraday, Maxwell and others before it took the proportions
that would lead Einstein and Infeld to say ``A new concept appears in physics, the most important invention since
Newton's time: the field.  It needed great scientific imagination to realize that it is not
the charges nor the particles, but the field in the space between the charges and the particles
which is essential for the description of physical phenomena.''  The question is: to deliver on
Schr\"{o}dinger's call to ``account for'' the phenomena of life, what new 
experiments and concepts  does expanding our domain of physical enquiry into the realm of the living demand? 

My reading of Schr\"{o}dinger's book is that he is arguing that, just
as the phenomena of heat and electrodynamics forced upon us
a myriad of new concepts such as temperature, entropy, the electric
field and radiation pressure, the phenomena of the living world similarly
demand the continued evolution of physics through new concepts.
In his section on
 ``New Laws to Be Expected in the Organism,'' 
 Schr\"{o}dinger reveals his hand:  ``What I wish to make
 clear in this last chapter is,
in short, that from all we have learnt about the
structure of living matter, we must be prepared to
find it working in a manner that cannot be
reduced to the ordinary laws of physics. 
And that
not on the ground that there is any ``new force'' or
what not, directing the behaviour of the single
atoms within a living organism, but because the
construction is different from  anything we have
yet tested in the physical laboratory.''  Indeed, would
anyone be surprised by the idea that, as we subject new
classes of phenomena to quantitative scrutiny in the physical
laboratory, yet again, as has been true every century since
the days of Tycho Brahe, new concepts will be demanded
of us?

 Schr\"{o}dinger makes this point beautifully through a  colorful analogy for thinking
about how the elements of the periodic table can be exploited
to totally different ends in living matter.
``To put it
crudely, an engineer, familiar with heat engines
only, will, after inspecting the construction of an
electric motor, be prepared to find it working
along principles which he does not yet
understand. He finds the copper familiar to him
in kettles used here in the form of long wires
wound in coils; the iron familiar to him in levers
and bars and steam cylinders here filling the
interior of those coils of copper wire. He will be
convinced that it is the same copper and the same
iron, subject to the same laws of Nature, and he
is right in that. The difference in construction is
enough to prepare him for an entirely different
way of functioning. He will not suspect that an
electric motor is driven by a ghost because it is
set spinning by the turn of a switch, without
boiler and steam.''

Schr\"{o}dinger thus concludes
``it needs no poetical imagination but only clear and sober scientific reflection
to recognize that we are here obviously faced with events whose regular and lawful
unfolding is guided by a `mechanism' entirely different from the `probability mechanism'
of physics.''
One category of ``new laws'' that might prove particularly potent in biology and
that have been given short shrift in the molecular biology era are 
strictly phenomenological models that make no reference to
underlying ``mechanism.'' 
These kinds of laws have formed a centerpiece of physics for more than three hundred
years in contexts ranging from the laws of elasticity to hydrodynamics.  Here what I have in mind is a  phenomenological link in mathematical terms that connects the variables that have
emerged from measurements.   Both the familiar Hooke's law and ideal gas law originally emerged
as highly powerful, and yet, strictly phenomenological reflections of what had been
measured in the laboratory.
Examples have already started to emerge in physical biology as well.  Recent high-resolution measurements make it possible to explore the growth of bacteria resulting in a number of propositions for bacterial  growth laws~\cite{2014-iyer-biswas-pnas, Jun2018,Harris2018}.  With no reference to the molecular underpinnings, these phenomenological laws engender corresponding phenomenological hypotheses for how populations of bacterial cells maintain a narrow distribution of cell sizes,  such as 
adder and timer models in which the growing cell either waits to divide until it has added a certain fixed quantity of material or until a fixed amount of time has elapsed~\cite{Campos2014}.  Another provocative
and beautiful set of phenomenological  laws has emerged concerning
the character  of the entire proteome~\cite{Schaechter1958, Scott2010}.  Here the idea
is that the growth rate of cells serves as a kind of ``state variable'' and that depending upon
this growth rate, the fraction of the proteome devoted to protein production (i.e., ribosomes) 
takes a very specific value.   Schr\"{o}dinger's chapter 6 asks us to think about
the emergence of order from order as seen in Figure~\ref{fig:NegativeEntropy}, a topic that has become central to modern
physical biology, whether in the context of bird flocks~\cite{Toner1998,Cavagna2014}  or cytoskeleton-motor systems such as those that partition chromosomes during cell division~\cite{Ramaswamy2010, Marchetti2013, Prost2015, Needleman2017}.
In all of these cases, phenomenological laws have  emerged and we are now
in a stage of science when the laws are tested and refined, the implications are further
examined and efforts are set forth to try and derive those laws from some deeper understanding
of the underlying processes.

Some of the many other areas in which I suspect we will find ``new physics'' focus on the ways in
which biological systems locally defy the tendency towards equilibrium.  Examples include a series of
problems such as the accuracy problem (how biological systems achieve such high
fidelity in comparison with what is implied by the ``statistical structure'' of classical physics),
the adaptation problem (how biological systems change their physico-chemical behavior
in real time in response to environmental cues), the reproducibility problem (how living organisms
achieve the same outcome such as the human body plan over and over again in a nearly error-free fashion), the structure problem (how structures such as the outer segment of photoreceptors
are constructed and maintained in the face of the noisy world that surrounds them).
I also suspect that there are  surprises before us in the unfinished business of dynamics, a subject that started
with the great successes  of classical
mechanics, then to the incomplete promise of thermodynamics and now the full-fledged challenge
of the out-of-equilibrium aspects of biological dynamics, whether the separation of chromosomes or the origin of species.
Of course, these ``problems'' are offered tentatively and subjectively
since every generation has to decide what are its most pressing problems, but my main point
is about the kind of solutions that we should demand in accounting for
biological phenomena.  

\section{A Manifesto:  Schr\"{o}dinger's Unfinished Business}

A perennial question for any scientist is: what to work on? Of course, there is no one
right answer and one of my own favorite answers is: whether young or old, scientists should work on whatever they are truly most
curious about understanding from the vast array of mysteries presented by the world around us.
In academia, much debate and angst are powered by this question in disguise
as we ask ourselves, should this person be hired or that grant be funded?
One of the ways that  people try to sharpen that question is by asking
whether we will  find new physics or new  biology, depending  upon within which
department that question is considered.    For a very thoughtful modern reflection on the question of
new physics more broadly by a noted physicist who worked in many domains of physics, see ``Does Astronomy Need `New Physics'?''~\cite{Ginzburg2001}.
Stated most succinctly, Schr\"{o}dinger's  short work ventures the 
guess that indeed, by looking at living matter, those who subscribe 
to the definition of understanding demanded in physics will find
new physics there.

As already alluded to throughout the article, the last 30 years have seen enormous progress at the interface
between physics and biology.   As for whether or not there is new physics to report, I think
it depends on how we take that question.  If we are practicing what Thomas Kuhn referred
to as ``normal science,'' there is no doubt that there has been an impressive array of 
results that surely count as new physics.  Examples abound,  whether in the context of cellular motility, detection and adaptation in
the context of chemotaxis~\cite{Purcell1977,Berg1977,Sourjik2002}, or in the analysis of
population genetics with its beautiful analogies between the Boltzmann distribution of
statistical mechanics and the distribution of allele frequencies~\cite{Lassig2007}, or in the surprising new features 
revealed in the study of active matter~\cite{Toner2018}.   In an  excellent series
of lectures, John Toner notes: ``...the biggest surprise in the entire field of active matter is that a `polar ordered dry active fluid phase' is even possible in two dimensions,''
 technical words behind
a fun and interesting example of the new physics that has emerged in thinking about bird flocks.
Yet another example is offered by the phenomenon of cytoplasmic streaming~\cite{Corti1774}, observed before
the founding of the United States, and yet only in the last decade has the natural language
of dimensionless variables allowed us to compare the relative importance of diffusion
and flow in transporting materials within large cells and to compute the kinds of flow patterns that emerge~\cite{Goldstein2008,Mayer2010,Goldstein2015,Munster2019}.  All of these examples constitute the activities of normal science and the act of
``doing physics'' on them has resulted in not only sharpening our questions, but also in
increasing the depth of our understanding.
In a playful turn of the millennium piece entitled ``Molecular Vitalism,'' Kirschner, Gerhart, and  Mitchison show how many of
the issues raised by Schr\"{o}dinger about what gives living organisms
their distinct physicochemical attributes remain as fresh now as they were in the 1940s~\cite{Kirschner2000}.  
``We do not question the importance of
genetics, nor dispute the role of DNA as the blueprint
for all the components of living systems, but we think
it worth asking to what extent the �postgenomic� view
of modern biology would convince a nineteenth century
vitalist that the nature of life was now understood.''

If we adopt a more sweeping view in which
we ask for a revolution in physics that has come on the heels of investigating 
biological phenomena, we may need to exercise a little more patience.   We shouldn't be surprised
by the glacial pace at which we get our revolutions.  There was a century between the initial
discovery of spectral lines and properly ``accounting for'' them in the series of classic
papers by  Schr\"{o}dinger~\cite{Schrodinger1978}.  That said,
I am wary of the common attitude that, because something has not been done, or worse
yet, because a given commentator themself has not done it, that means that it cannot be done. 
I suspect that in the physics pedagogy a hundred years hence (if humans are still teaching each other important ideas by then), there will be courses whose central ambition will be
to explain the harvest of the revolution that resulted from physicists trying to answer the
question  ``what is life?'' to their own satisfaction, in much the same way that we have physics courses
dedicated to the question of ``what is matter?''

Despite both the insight and promise of Schr\"{o}dinger's thinking in ``What is Life?'', 
not all responses to his book were positive~\cite{Dronamraju1999}.  
Three greatly accomplished scientists who came down on the negative side of the ledger were famed American geneticist HJ Muller, Caltech chemist and visionary Linus Pauling, and the structural biologist Max Perutz,  Schr\"{o}dinger's compatriot.  Their reactions strike to the very heart of how different fields view what questions are interesting and what constitutes acceptable
answers to those questions~\cite{Keller2002}.   Indeed, as is now clear, a major thrust of this essay has been that in adopting   Schr\"{o}dinger's physicist definition of 
 understanding, we will see that  the study of living matter will demand new physics.  
 Though differences in
philosophy about what it means to understand something explain some of
the negative reaction to Schr\"{o}dinger's classic, I find a marked lack of generosity given the circumstances of the book's origin as a written summary of public lectures.  

In  a centenary volume celebrating
Schr\"{o}dinger's contributions to modern science~\cite{Kilmister1989}, Pauling wrote  ``Schr\"{o}dinger's discussion of thermodynamics is vague and superficial to an extent that
should not be tolerated even in a popular lecture.''  
Writing in the same volume, 
Perutz is even more scathing, making comments such as
``Sadly, however, a close study of his book and of the related literature has shown me that what
was true in his book was not original, and most of what was original was known not to be true even
when the book was written....the apparent contradiction between life and the statistical laws of
physics can be resolved by invoking a science largely ignored
by Schr\"{o}dinger.  That science is chemistry.''
Hence, as did Crick, Perutz takes Schr\"{o}dinger to task for ignoring ``chemistry,'' which
he argues would make the stability of genetic information clear.
Here I  part  ways with Perutz since as our ability to routinely melt DNA in our PCR machines shows, the stability of the genetic material
is intimately related to precisely the thermal physics discussed by Schr\"{o}dinger.  Understanding
the high-fidelity of the processes of the central dogma, to name but one example that falls outside
the purview of both classical statistical physics and the chemistry of the mid 20$^{th}$ century,  demands much more of us
than Boltzmann distributions and chemical bonding.

It is intriguing to explore the  claim
that if Schr\"{o}dinger had but only appealed to the science of chemistry, no mysteries would have
remained for the classical physicist trying to
``account for'' the living organism.  Having spent now nearly a decade trying to come to terms with the field that
Perutz was central in creating, namely, the subject of allostery, I remain more skeptical
than ever of what I will call the salt-bridge argument~\cite{Perutz1978, Perutz1998}, an example of the conviction that if we only understand
the atomic-level structures of the macromolecules of the living world, then ``mechanism'' and understanding will unfold
before our eyes. 
 Really, what we are talking about here is precisely the kind of polarizing debate
that separates our political lives, what Thomas Sowell christened a ``conflict of visions.''  
Schr\"{o}dinger was not invalidating or critiquing the world view of chemists or biologists, he
was trying to explain what it looks like for physicists to ``account for'' a subject.
His views are echoed forcefully and eloquently by today's leading biophysical thinkers~\cite{Bialek2012, Nelson2015, Goldstein2018, Bialek2015}.
The point of my loving review of  Schr\"{o}dinger's little book ``What is Life?'' 75 years on is that,
to really understand the book's meaning, we have to remember both the question being
considered and the audience being addressed.   In my view, it is a mistake to think of his work  as a manifesto about biology for biologists. 
 It is a manifesto about the frontiers of physics and the way that every time physics tackles
 new classes of phenomena, it requires new concepts and ultimately results in
 the formulation of new laws.
It is also a manifesto about the unity of nature.
Nature cares not for the names of our subjects.  Names such as physics
and biology are a strictly human
conceit and the understanding of the phenomenon of life might require us
to blur the boundaries between these fields.

Sidney Brenner once quipped that, in research, one
should either be 6 months ahead of the scientific pack or 30 years behind.  There is much
to that remark since over and over again, pathbreaking discoveries are made when
new technologies are used to reconsider ``old'' problems.  Nowhere is this more true 
than in the case of the hydrogen atom, one of the most remarkable gifts ever
given to science~\cite{Rigden2002}.  Hydrogen has served as the quintessential test case
for what it means to ``account for'' the behavior of atoms (spectral lines), the nucleus (the deuteron),
the coupling between radiation and matter, Bose-Einstein condensates and beyond.
Similar case studies are only waiting to be exploited in biology once Schr\"{o}dinger's notion
of what it means to account for a phenomenon is accepted.  The study of living matter needs
its hydrogen atoms.
Erwin Schr\"{o}dinger's remarkable
`What is Life?' makes it clear that  Brenner could have gone even  farther and exhorted
us to search 75 years into the past to find an inspiring charge for the future.  \\

\noindent {\bf Acknowledgments}\\

I am especially grateful to David Booth who kindly suggested that I undertake
this labor of love,
and to Christina Hueschen with whom I have repeatedly discussed
Schr\"{o}dinger's classic book over the last several years.  
My biggest debt in learning about the confluence
of biology and physics is to 
 my book  co-authors Jane Kondev, Julie  Theriot, Hernan Garcia, Christina  Hueschen,
Ron Milo, Wallace Marshall and Thomas Lecuit.  In addition, I have learned so much about physics, biology and their  intersection from
 a veritable who's who of  deep thinkers on the state of the art
in modern science. For either their direct or indirect help with this article, I want to especially thank  Clarice Aiello, Howard Berg, Bill Bialek, Curt Callan, Anders Carlsson, Griffin Chure, Ken Dill, Ethan Garner, Bill Gelbart, Lea Goentoro, Ray Goldstein, Stephan Grill, Christoph Haselwandter, Hopi Hoekstra, Joe Howard, Tony Hyman, Sri Iyer-Biswas, Quincey Justman, Marc Kirschner, Heun Jin Lee, Jennifer Lippincott-Schwartz, Niko McCarty, Madhav Mani, Tim Mitchison, Chris Miller, Andrew Murray, Phil Nelson, Hirosi Ooguri, Mariela Petkova, Molly Phillips, Steve Quake,  Manuel Razo, Udo Seifert, Pierre Sens, Lubert Stryer, Mark Uline, Ron Vale, Aleks Walczak, Jon Widom, Chris Wiggins, Ned Wingreen and Carl Zimmer. I am grateful to all of these impressive thinkers for useful discussions and/or commenting on the manuscript, though the views expressed here should not be blamed on them.  I am also grateful to the NIH for support through award numbers DP1OD000217 (Director's Pioneer Award) and R01 GM085286.  The trust and financial support of this great institution make it possible for today's scientists to grapple with the endless fascination of trying to answer Schr\"{o}dinger's classic question,  ``What is Life?''
\\

\clearpage

\noindent{\bf References}\\

\bibliography{PaperLibrarySchrodinger}

\begin{thebibliography}{10}

\bibitem{Schrodinger1992}
E.~Schr\"{o}dinger.
\newblock {\em What is Life? with Mind and Matter \& Autobiographical
  Sketches}.
\newblock Cambridge University Press, Cambridge, England, 1992.

\bibitem{Dronamraju1999}
KR~Dronamraju.
\newblock {Erwin Schr\"{o}dinger and the Origins of Molecular Biology}.
\newblock {\em Genetics}, 153:1071--1076, 1999.

\bibitem{Moore2015}
Walter Moore.
\newblock {\em {Schr\"{o}dinger Life and Thought}}.
\newblock Cambridge University Press, Cambridge: England, 2015.

\bibitem{Gribbin2013}
John Gribbin.
\newblock {\em {Erwin Schr\"{o}dinger and the Quantum Revolution}}.
\newblock John Wiley \& Sons, New Jersey, 2013.

\bibitem{Judson1996}
Horace~Freeland Judson.
\newblock {\em {The Eighth Day of Creation}}.
\newblock Cold Spring Harbor Laboratory Press, New York, 1996.

\bibitem{Harrison1939}
G.R. Harrison.
\newblock {\em Massachusetts Institute of Technology Wavelength Tables with
  Intensities in Arc, Spark or Discharge Tube of More Than 100,000 Spectrum
  Lines}.
\newblock John Wiley and Sons, Inc., New York, New York, 1939.

\bibitem{Mehra1982}
J~Mehra and Rechenberg.
\newblock {\em {The Historical Development of Quantum Theory, Vol. 1, Part 1}}.
\newblock Springer Verlag, New York, 1982.

\bibitem{Planck1902}
M.~Planck.
\newblock {\"{U}ber die Natur des weisen Lichtes}.
\newblock {\em Annalen der Physik}, 7:390--400, 1902.

\bibitem{Klein1973}
M~Klein.
\newblock {Mechanical Explanation at the End of the Nineteenth Century}.
\newblock {\em Centaurus}, 17:58--82, 1973.

\bibitem{Brush1986}
SG~Brush.
\newblock {\em {The Kind of Motion We Call Heat}}.
\newblock North Holland, Amsterdam, 1986.

\bibitem{Pais1979}
A.~Pais.
\newblock {Einstein and the Quantum Theory}.
\newblock {\em Rev. Mod. Phys.}, 51:863--914, 1979.

\bibitem{Turing1952}
A.~M. Turing.
\newblock {The Chemical Basis of Morphogenesis}.
\newblock {\em Philosophical Transactions of the Royal Society of London Series
  B-Biological Sciences}, 237(641):37--72, 1952.

\bibitem{Hopfield1974}
J.~J. Hopfield.
\newblock {Kinetic proofreading: a new mechanism for reducing errors in
  biosynthetic processes requiring high specificity}.
\newblock {\em Proc Natl Acad Sci U S A}, 71(10):4135--9, 1974.

\bibitem{Ninio1975}
J.~Ninio.
\newblock {Kinetic amplification of enzyme discrimination}.
\newblock {\em Biochimie}, 57(5):587--95, 1975.

\bibitem{Berg1977}
H.~C. Berg and E.~M. Purcell.
\newblock Physics of chemoreception.
\newblock {\em Biophys J}, 20(2):193--219, 1977.

\bibitem{Toner2018}
J.~Toner.
\newblock {Why walking is easier than pointing: Hydrodynamics of dry active
  matter}.
\newblock {\em arXiv}, 1812.00310v1, 2018.

\bibitem{Pauling1957}
L.~Pauling.
\newblock {The probability of errors in the process of synthesis of protein
  molecules in Festschrift fuer Pr. Dr. Arthur Stoll }.
\newblock 1957.

\bibitem{Bialek2012}
William~S. Bialek.
\newblock {\em Biophysics : searching for principles}.
\newblock Princeton University Press, Princeton, NJ, 2012.

\bibitem{Savage2010}
D.~F. Savage, B.~Afonso, A.~H. Chen, and P.~A. Silver.
\newblock {Spatially ordered dynamics of the bacterial carbon fixation
  machinery}.
\newblock {\em Science}, 327(5970):1258--61, 2010.

\bibitem{Rosenfeld2005}
N.~Rosenfeld, J.~W. Young, U.~Alon, P.~S. Swain, and M.~B. Elowitz.
\newblock {Gene regulation at the single-cell level}.
\newblock {\em Science}, 307(5717):1962--5, 2005.

\bibitem{Rosenfeld2006}
N.~Rosenfeld, T.~J. Perkins, U.~Alon, M.~B. Elowitz, and P.~S. Swain.
\newblock {A fluctuation method to quantify in vivo fluorescence data}.
\newblock {\em Biophys J}, 91(2):759--66, 2006.

\bibitem{Seifert2012}
U.~Seifert.
\newblock {Stochastic thermodynamics, fluctuation theorems and molecular
  machines}.
\newblock {\em Reports on Progress in Physics}, 75:126001, 2012.

\bibitem{Mahajan2010}
Sanjoy Mahajan.
\newblock {\em Street-fighting mathematics : the art of educated guessing and
  opportunistic problem solving}.
\newblock MIT Press, Cambridge, Mass., 2010.

\bibitem{SchwartzBook}
D.~N. Schwartz.
\newblock {\em {The Last Man Who Knew Everything}}.
\newblock Basic Books, New York, New York, 2017.

\bibitem{Sloan2011}
P.~R. Sloan and B.~Fogel.
\newblock {\em Creating a Physical Biology The Three-Man Paper and Early
  Molecular Biology}.
\newblock University of Chicago Press, Chicago, Illinois, 2011.

\bibitem{Perutz1987}
M~Perutz.
\newblock {Physics and the riddle of life}.
\newblock {\em Nature}, 326:555--559, 1987.

\bibitem{Tanford2004}
Charles Tanford.
\newblock {\em {Ben Franklin Stilled the Waves}}.
\newblock Oxford University Press, New York, 2004.

\bibitem{Langmuir1917}
Irving Langmuir.
\newblock {The constitution and fundamental properties of solids and liquids.
  II. Liquids.}
\newblock {\em Journal of the American Chemical Society}, 39:1848--1906, 1917.

\bibitem{Einstein1938}
A.~Einstein and L.~Infeld.
\newblock {\em {The Evolution of Physics}}.
\newblock Cambridge University Press, Cambridge, England, 1938.

\bibitem{2014-iyer-biswas-pnas}
Srividya Iyer-Biswas, Charles~S. Wright, Jonathan~T. Henry, Klevin Lo,
  Stanislav Burov, Yihan Lin, Gavin~E. Crooks, Sean Crosson, Aaron~R. Dinner,
  and Norbert~F. Scherer.
\newblock Scaling laws governing stochastic growth and division of single
  bacterial cells.
\newblock {\em Proc. Natl. Acad. Sci. U.S.A.}, 111(45):15912--15917, Nov 2014.

\bibitem{Jun2018}
S.~Jun, F.~Si, R.~Pugatch, and M.~Scott.
\newblock {Fundamental principles in bacterial physiology - history, recent
  progress, and the future with focus on cell size control: a review}.
\newblock {\em Rept. Prog. Phys.}, 81(5):056601, 2018.

\bibitem{Harris2018}
L.~K. Harris and J.~A. Theriot.
\newblock {Surface Area to Volume Ratio: A Natural Variable for Bacterial
  Morphogenesis}.
\newblock {\em Trends Microbiol}, 26(10):815--832, 2018.

\bibitem{Campos2014}
M.~Campos, I.~V. Surovtsev, S.~Kato, A.~Paintdakhi, B.~Beltran, S.~E. Ebmeier,
  and C.~Jacobs-Wagner.
\newblock {A constant size extension drives bacterial cell size homeostasis}.
\newblock {\em Cell}, 159(6):1433--46, 2014.

\bibitem{Schaechter1958}
M.~Schaechter, O.~Maaloe, and N.~O. Kjeldgaard.
\newblock {Dependency on medium and temperature of cell size and chemical
  composition during balanced grown of {\it Salmonella typhimurium}}.
\newblock {\em J Gen Microbiol}, 19(3):592--606, 1958.

\bibitem{Scott2010}
M.~Scott, C.~W. Gunderson, E.~M. Mateescu, Z.~Zhang, and T.~Hwa.
\newblock Interdependence of cell growth and gene expression: origins and
  consequences.
\newblock {\em Science}, 330(6007):1099--102, 2010.

\bibitem{Toner1998}
J.~Toner and Y.~Tu.
\newblock {Flocks, herds, and schools: A quantitative theory of flocking}.
\newblock {\em Phys. Rev. E.}, 58:4828--4858, 1998.

\bibitem{Cavagna2014}
A.~Cavagna and I.~Giardina.
\newblock {Bird Flocks as Condensed Matter}.
\newblock {\em Annu. Rev. Condens. Matter Phys.,}, 5:183--207, 2014.

\bibitem{Ramaswamy2010}
S.~Ramaswamy.
\newblock {The Mechanics and Statistics of Active Matter}.
\newblock {\em Annu. Rev. Condens.Matter Phys.}, 1:323--345, 2010.

\bibitem{Marchetti2013}
M.~C. Marchetti, J.~F. Joanny, S.~Ramaswamy, T.~B. Liverpool, J.~Prost, M.~Rao,
  and R.~Aditi Simha.
\newblock {Hydrodynamics of soft active matter}.
\newblock {\em Rev. Mod. Phys.}, 85:1143--1189, 2013.

\bibitem{Prost2015}
J.~Prost, F.~Julicher, and J.~F. Joanny.
\newblock {Active gel physics}.
\newblock {\em Nature Physics}, 11(2):111--117, 2015.

\bibitem{Needleman2017}
Daniel Needleman and Zvonimir Dogic.
\newblock Active matter at the interface between materials science and cell
  biology.
\newblock {\em Nature Reviews Materials}, 2:17048 EP, Jul 2017.

\bibitem{Ginzburg2001}
V.~L. Ginzburg.
\newblock {\em {The Physics of a Lifetime}}.
\newblock Springer-Verlag, Berlin, 2001.

\bibitem{Purcell1977}
EM~Purcell.
\newblock {Life at Low Reynolds Number}.
\newblock {\em {American Journal of Physics}}, {45}({1}):{3--11}, {1977}.

\bibitem{Sourjik2002}
V.~Sourjik and H.~C. Berg.
\newblock {Receptor sensitivity in bacterial chemotaxis}.
\newblock {\em Proc Natl Acad Sci U S A}, 99(1):123--7, 2002.

\bibitem{Lassig2007}
M.~Lassig.
\newblock {From biophysics to evolutionary genetics: statistical aspects of
  gene regulation}.
\newblock {\em BMC Bioinformatics}, 8 Suppl 6:S7, 2007.

\bibitem{Corti1774}
B.~Corti.
\newblock {\em Osservazioni microscopiche sulla tremella e sulla circolazione
  del fluido in una pianta acquajuola}.
\newblock Rocchi, 1774.

\bibitem{Goldstein2008}
R.~E. Goldstein, I.~Tuval, and J.~W. van~de Meent.
\newblock {Microfluidics of cytoplasmic streaming and its implications for
  intracellular transport}.
\newblock {\em Proc Natl Acad Sci U S A}, 105(10):3663--7, 2008.

\bibitem{Mayer2010}
M.~Mayer, M.~Depken, J.~S. Bois, F.~Julicher, and S.~W. Grill.
\newblock Anisotropies in cortical tension reveal the physical basis of
  polarizing cortical flows.
\newblock {\em Nature}, 467(7315):617--U150, 2010.

\bibitem{Goldstein2015}
R.~E. Goldstein and J.~W. van~de Meent.
\newblock {A physical perspective on cytoplasmic streaming}.
\newblock {\em Interface Focus}, 5(4):20150030, 2015.

\bibitem{Munster2019}
S.~Munster, A.~Jain, A.~Mietke, A.~Pavlopoulos, S.~W. Grill, and P.~Tomancak.
\newblock {Attachment of the blastoderm to the vitelline envelope affects
  gastrulation of insects}.
\newblock {\em Nature}, 568(7752):395--399, 2019.

\bibitem{Kirschner2000}
M.~Kirschner, J.~Gerhart, and T.~Mitchison.
\newblock {Molecular "vitalism"}.
\newblock {\em Cell}, 100(1):79--88, 2000.

\bibitem{Schrodinger1978}
E.~Schr\"{o}dinger.
\newblock {\em {Collected Papers on Wave Mechanics}}.
\newblock Chelsea Publishing Company, New York, 1978.

\bibitem{Keller2002}
Evelyn~Fox Keller.
\newblock {\em Making Sense of Life}.
\newblock Harvard University Press, Cambridge, Massachusetts, 2002.

\bibitem{Kilmister1989}
CW~Kilmister~(editor).
\newblock {\em {Schrodinger: Centenary Celebration of a Polymath}}.
\newblock Cambridge University Press, Cambridge, 1989.

\bibitem{Perutz1978}
M.~F. Perutz.
\newblock {Electrostatic effects in proteins}.
\newblock {\em Science}, 201(4362):1187--91, 1978.

\bibitem{Perutz1998}
S.~Bettati, A.~Mozzarelli, and M.~F. Perutz.
\newblock {Allosteric mechanism of haemoglobin: rupture of salt-bridges raises
  the oxygen affinity of the T-structure}.
\newblock {\em J Mol Biol}, 281(4):581--5, 1998.

\bibitem{Nelson2015}
Philip~Charles Nelson, Sarina Bromberg, Ann Hermundstad, and Jason Prentice.
\newblock {\em Physical models of living systems}.
\newblock W.H. Freeman \& Company, New York, NY, 2015.

\bibitem{Goldstein2018}
R.~E. Goldstein.
\newblock Are theoretical results 'results'?
\newblock {\em Elife}, 7, 2018.

\bibitem{Bialek2015}
W.~Bialek.
\newblock {Perspectives on theory at the interface of physics and biology}.
\newblock {\em arXiv}, 1512.08954v1, 2015.

\bibitem{Rigden2002}
J.~S. Rigden.
\newblock {\em {Hydrogen The Essential Element}}.
\newblock Harvard University Press, Cambridge, Mass, 2002.

\end{thebibliography}
\bibliographystyle{unsrt}

\end{document}